\documentclass[prb,pre,aps,twocolumn,amsmath,amssymb,floatfix,
superscriptaddress]{revtex4-1}
\usepackage[dvips]{graphics}
\usepackage{graphicx}
\usepackage{color}
\usepackage{blindtext}
\usepackage{amsmath}
\definecolor{dred}{rgb}{0.75,0,0}
\usepackage{graphicx}
\usepackage{dcolumn}
\usepackage{bm}
\usepackage{xcolor}
\usepackage{braket}
\usepackage{physics}
\usepackage{appendix}
\usepackage{hyperref}
\hypersetup{
 colorlinks=true,
citecolor=magenta,filecolor=blue,
 linkcolor=magenta,
 }

\usepackage{graphicx}
\usepackage{dcolumn}
\usepackage{bm}
\usepackage{xcolor}

\definecolor{codegreen}{rgb}{0,0.6,0}
\definecolor{codegray}{rgb}{0.5,0.5,0.5}
\definecolor{codepurple}{rgb}{0.58,0,0.82}
\definecolor{backcolour}{rgb}{0.95,0.95,0.92}
 
\begin{document}

\preprint{APS/123-QED}

\title{\textcolor{blue}{Complete escape from localization on a hierarchical lattice: A Koch fractal with all states extended}} 

\author{Sougata Biswas}
\affiliation{Department of Physics, Presidency University, 86/1 College Street, Kolkata, West Bengal - 700 073, India}
\affiliation{sougata.rs@presiuniv.ac.in}
\author{Arunava Chakrabarti}
\affiliation{Department of Physics, Presidency University, 86/1 College Street, Kolkata, West Bengal - 700 073, India}
\affiliation{arunava.physics@presiuniv.ac.in}
\date{\today}

\begin{abstract}
An infinitely large Koch fractal is shown to be capable of sustaining {\it only} extended, Bloch-like eigenstates, if certain parameters of the Hamiltonian describing the lattice are numerically correlated in a special way, and a magnetic flux of a special strength is trapped in every loop of the geometry. We describe the system within a tight binding formalism and prescribe the desired correlation between the numerical values of the nearest neighbor overlap integrals, along with a special value of the magnetic flux trapped in the triangular loops decorating the fractal. With such conditions, the lattice, despite the absence of translational order of any kind whatsoever, yields an absolutely continuous eigenvalue spectrum, and becomes completely transparent to an incoming electron with any energy within the allowed band. The results are analytically exact. An in-depth numerical study of the inverse participation ratio and the two-terminal transmission coefficient corroborates our findings. Our conclusions remain valid for a large set of lattice models, built with the same structural units, but beyond the specific geometry of a Koch fractal, unraveling a subtle universality in a variety of such low dimensional systems.
\end{abstract}

\maketitle

\section{Introduction}
\label{intro}
Interference between multiple scattering events leads to a localization of waves in a disordered environment. This is the essence of what we now know as Anderson localization~\cite{anderson,tvr}. The details of the phenomenon are of course sensitive to the dimensionality of the system, and tight binding studies of lattice models in three, two, and one dimensions reveal that, while there is a critical concentration of the disorder in three dimensions to see the absence of diffusion~\cite{anderson}, in two dimensions~\cite{abrahams} and in one dimension~\cite{borland,deylon} all single-particle states are localized in general, for any arbitrary amount of disorder.
This disorder-driven quantum interference effect, conceptualised more than sixty years ago, is of everlasting interest as the manifestation of disorder is ubiquitious~\cite{kramer,lagen} and is experimentally observed in recent past for a wide variety of systems using light~\cite{albada,wiersma,martin,segev}, Bose-Einstein condensates~\cite{roati}, or the very recent one involving cold atoms~\cite{white}, to name a few. 

A substantial volume of the existing literature addresses different aspects of disorder-induced localization~\cite{economou1, herbert,thouless,economou2,anderson2,guinea,rudo0,francisco,sacha} and this field of research has been further enriched by the discovery of localization even in the absence of disorder, viz, in translationally invariant lattice geometries where local lattice topology becomes the root cause of localization~\cite{sutherland}. This observation pioneered extensive research in the physics of the so-called {\it flat bands} and {\it compact localized states} in a class of `engineered' lattice models in quasi-one dimension, two or even in three dimensions~\cite{bodyfelt,ajith,sergej1,biplab,sergej2,rudo1,rudo2,rudo3}. 

In the present communication, we investigate the counterintuitive case of a {\it complete delocalization} of all the single-particle states, representing, say, an electron propagating in a disordered environment, where the disorder has a {\it structure}. Delocalization of single-particle states in lattices with geometrically correlated disorder is not new, and began with the pioneering `random dimer model' (RDM)~\cite{dunlap,wu}.  It was substantiated later by several other interesting works on (geometrically) correlated disordered lattices~\cite{goff,kundu,sanchez,heinrichs,grosso}, and also in a whole class of quasiperiodic lattices where the {\it dimer-like}~\cite{dunlap} geometric correlation is built in the growth sequences~\cite{arunava1,arunava2,arunava3,macia1,arunava4}. Delocalized eigenstates are also seen in deterministic fractal lattice models~\cite{arunava5,wang,arunava6}, and the origin of the existence of extended, delocalized single-particle states in such lattices can be traced back to the lattice geometry as a whole, in contrast to {\it local} clusters of sites, as shown in the RDM's or the quasiperiodic lattices referred to above.


\begin{figure}[ht]
\centering
\includegraphics[width=\columnwidth]{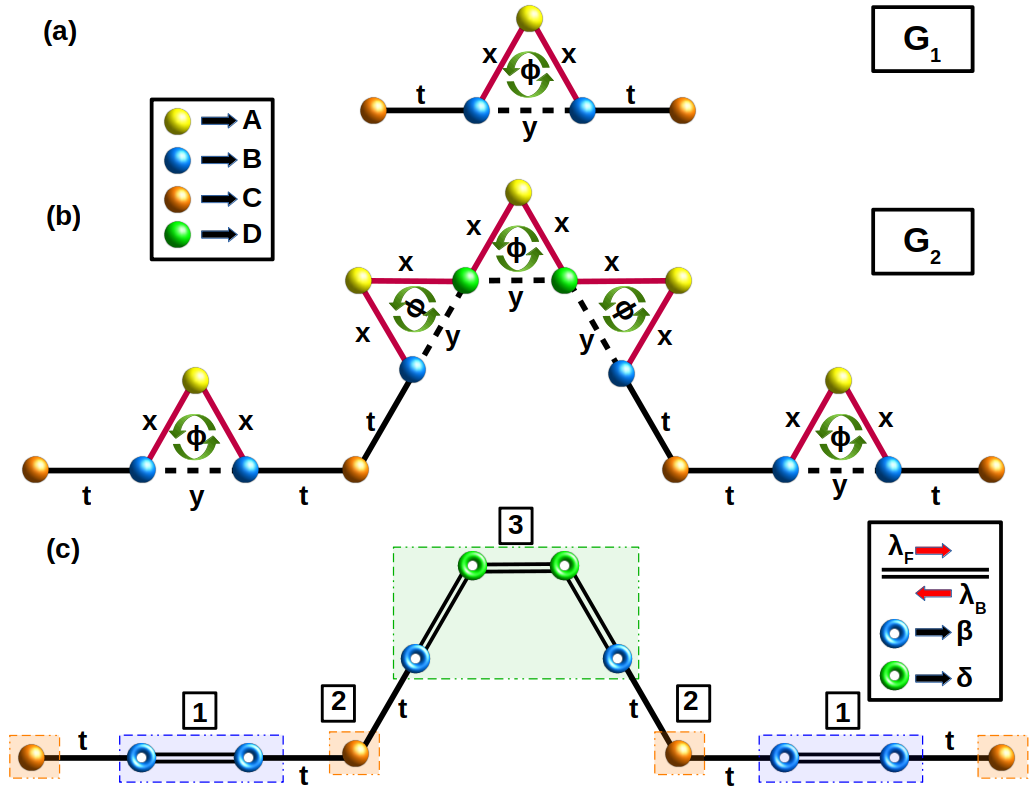}
\caption{(Color online) (a) The building block (seed) of a Koch fractal. (b) The Koch fractal is in its second generation. The variety of sites $A$, $B$, $C$, and $D$ are explained through the color code in the inset. (c) A linear chain is obtained from (b) after decimating out the $A$ sites. The magnetic flux trapped in the triangular plaquettes, in each figure, is shown as $\Phi$. }  
\label{koch-lattice}
\end{figure}

In all the previous works mentioned above, delocalization (leading to extended eigenfunctions) is observed only at {\it special}  energy eigenvalues. In an infinitely large quasiperiodic lattice~\cite{arunava1,arunava2,arunava3} or in a deterministic fractal in its thermodynamic limit~\cite{wang,arunava6}, one can extract an infinity of such {\it resonant} energy eigenvalues leading to unscattered single-particle states. But, the distribution of such energy values is always of a discrete nature, and the energies do not really form a continuous band, and definitely do not span the entire spectrum of eigenvalues.

In this backdrop, it has recently been proposed and shown that a class of one-dimensional topologically disordered lattices can indeed have a major part of the energy spectrum, or even the entire spectrum in some cases, populated by extended Bloch-like eigenstates only~\cite{skm,bpac,anbpac,amrita}. The entire spectrum, or most of it, depending upon the geometry of the lattice, becomes absolutely continuous. Such lattices are linear, periodic, and have side-coupled atomic clusters. Typically the systems are described by a tight binding Hamiltonian with {\it off-diagonal} disorder. A very nominal quasi-one dimensionality is introduced through the coupling of the clusters of atomic sites from one side of the periodic backbone. This is crucial. One then finds that a special correlation between the {\it numerical values} of the nearest neighbor hopping (overlap) integrals is needed to observe the delocalization effect. The studies~\cite{skm,bpac,anbpac,amrita} based on the solution of time-independent Schr\"{o}dinger equation was later substantiated by a study of quantum dynamics of similar systems~\cite{arka}.
In the present paper, we look deeper into this aspect of the problem, that is, we look into the possibility and the mechanism of generating an absolutely continuous band of energy eigenvalues with {\it all} states extended, in a disordered environment. We choose a recursively grown Koch fractal (KF) geometry, a well studied example of a hierarchical lattice~\cite{andrade1,macia,andrade2,zheng,zhao,achiam,zhang} as our system of interest. The KF lattice is deterministic, but lacks translational invariance. Generating extended eigenstates is such lattices, especially a continuous band of the same, is very much non-trivial. The KF geometry considered here is just an example, and towards the end of the communication, we explain that the conclusion drawn in this paper remains valid for many more lattice varieties, other than that considered here. 

The KF geometry in this work (Fig.~\ref{koch-lattice}(a, b)) is described using a tight binding Hamiltonian with a distribution of the nearest neighbor hopping integrals. Due to the hierarchical nature of the growth of the fractal, the constituent clusters, viz, a {\it triangle} and a {\it dot}, are non-periodically placed along the backbone of the lattice. There is no translational invariance, and the system, in general, has poor electrical conductance. The self-similarity of the geometry facilitates the application of the real space renormalization group (RSRG) methods to study the eigenvalue spectrum, and this has been exploited~\cite{arunava-koch1,arunava-koch2} to extract the density of states and also to identify clusters of isolated {\it delocalized} (extended, but non-Bloch) single-particle states at special, discrete energy values. However, no report is available yet in regard to achieving, by any mechanism, a completely continuous band structure in such geometries. This is the central motivation of the present work. We solve this problem in an analytically exact way.

Our results are very interesting. The model is described in section II. We show that a subtle correlation between the {\it numerical values} of the nearest neighbor hopping integrals on a Koch curve geometry can render the entire energy spectrum {\it absolutely continuous} with {\it all} the eigenstates having en {\it extended} character. The entire fractal geometry, however large, becomes {\it completely transparent} to an incoming electron with any energy lying within the band. The energy band, under the special correlation between the hopping integrals, shows the gross features of a perfectly periodic one-dimensional lattice of identical atoms, with a typical singularity at the edges of the band. This is discussed in detail, through a display of the density of states and the transmission spectrum, and is laid out in section III. That the correlation is of a much more general nature, is discussed next in section IV, and in section V we draw the conclusion. The mathematical details, needed to appreciate the results, are provided in the Appendices.

\section{The model}
A Koch fractal lattice is recursively grown from a seed shown in Fig.~\ref{koch-lattice} (a). The lattice, grown to its second generation is shown in Fig.\ref{koch-lattice}(b), and the subsequent generations are easily understood~\cite{macia,arunava-koch1}. The four kinds of atomic sites (vertices) are distinguished depending on the number of the nearest neighbors they have, viz, $A$ (yellow vertex of a triangle), $B$ (blue, base atoms of a triangle), $C$ (orange colored sites with two nearest neighbors), and $D$ (green colored sites with four nearest neighbors), The colors are explained by the codes given in Fig.~\ref{koch-lattice}. There is a uniform magnetic field, piercing each triangle perpendicular to its plane. The field results in a  magnetic flux $\Phi$ `trapped' in every triangular plaquette~\cite{arunava-koch1}. This breaks the time-reversal symmetry only {\it locally}, that is when an electron circulates around the triangle. 

The system is described using a tight binding Hamiltonian, viz, 
\begin{equation} 
H = \epsilon \ket{i}\bra{i} + \sum_{<ij>} \tau_{ij} \ket{i}\bra{j} + h.c.
\label{hamiltonian}
\end{equation}
Here, $\epsilon$ is the uniform on-site potential (set equal to zero in all our numerical calculations), and $\tau_{ij}$ is the nearest neighbor hopping integral connecting the $i$-th site with its nearest $j$-th sites. We will assign a value $\tau_{ij}=t$, a constant, when the hopping takes place between the adjacent sites lying on the backbone and does not encircle the trapped flux (that is, when the hopping is not around a triangle).  $\tau_{ij} = x \exp (\pm i\theta_{x})$ when an electron hops along the angular sides of a triangle, and $\tau_{ij}=y \exp (\pm i\theta_y)$, for a hop along the `base' of a triangle. $\theta_x$ 
 and $\theta_y$ are the Peierls' phases picked up by a propagating electron only when it moves along an arm of a triangle and is related to the total flux $\Phi$ trapped in the triangle through the equation $2\theta_x+\theta_y=2 \pi \Phi /\Phi_0$. $\Phi_0=hc/e$ is the fundamental flux quantum. For simplicity, we have set all bond-lengths to be equal to unity. 

As a result of the broken time-reversal symmetry along the sides of a triangle, $\theta_{ji}=-\theta_{ij}$. Without losing any generality, we choose the positive sign for $\theta_x$ or $\theta_y$ when the electron `hops' anti-clockwise along an arm of the triangle. We refer to it as the `forward' hopping from now on. Naturally, the term `backward' hopping will be used when $\theta_{ij}$ will have a negative sign as the electron hops back in a clockwise direction. For hopping along the bond on the backbone between vertices $B$ and $C$, not enclosing any flux, $\theta_{ij}=0$, and thus $\tau_{ij}=t$.

The time-independent Schr\"{o}dinger equation $H\psi = E \psi$ is cast into an equivalent set of discrete difference equations. The equations for the atomic sites at the vertices of a triangle read,
\begin{equation} 
(E-\epsilon) \psi_i = \sum_j \tau_{ij}  \psi_j
\label{difference}
\end{equation}
with $\tau_{ij}=t$, $x \exp (\pm i \theta_x)$, or $y \exp (\pm i\theta_y)$ as the case may be, and the sign of the Peierls' phase chosen appropriately. The summation runs over the nearest neighbors of the vertex considered. 

In the next section, we will use the set of Eqs.~\eqref{difference} to explain our findings.

\section{The Commuting transfer matrices and the continuous spectrum}
\subsection{The correlation and the commutation} 
Using Eqs.~\eqref{difference} we first decimate out~\cite{decimation} the $A$-sites (yellow colored) throughout the KF geometry, and map the entire lattice onto a purely one-dimensional chain (Fig.~\ref{koch-lattice}(c)). The $B$ and the $D$ sites in the parent KF are now {\it renormalized} to $\beta$ and $\delta$ sites, and they have energy-dependent, renormalized values of the on-site potentials, viz,
\begin{eqnarray}
    \epsilon_\beta & = & \epsilon+\frac{x^2}{(E-\epsilon)} \nonumber \\
    \epsilon_\delta & = & \epsilon+\frac{2 x^2}{(E-\epsilon)}
    \label{beta-delta}
\end{eqnarray}
The decimation also renormalizes the hopping amplitudes across the $\beta\beta$, $\beta\delta$, and $\delta\delta$ pairs of sites on the effectively linear chain shown in Fig.~\ref{koch-lattice}(c). Owing to the broken time-reversal symmetry caused by the trapped flux $\Phi$, the {\it forward} ($F$) and {\it backward} ($B$)  hopping integrals across such pairs are complex conjugates of each other. The `forward' hopping integrals are given by, 
$t_{\beta\beta}^F = t_{\beta\delta}^F = t_{\delta\delta}^F = \lambda \exp (i\xi) \equiv \lambda_F$ (say), where, 
\begin{eqnarray}
    \lambda & = & \sqrt{y^2 + \frac{x^4}{(E-\epsilon)^2}+ 
    \frac{2 x^2y \cos(2\pi \Phi/\Phi_0)}{E-\epsilon}} \nonumber \\
    \xi & = & \tan^{-1} \left [ \frac{y \sin(\theta_y) -\frac{x^2 \sin(2 \theta_x)}{E-\epsilon}} {y \cos (\theta_y) +\frac{x^2 \cos(2 \theta_x)}{E-\epsilon}} \right ]
\end{eqnarray}
The `backward' hoppings, as expected, are $t_{\beta\beta}^B = t_{\beta\delta}^B = t_{\delta\delta}^B = \lambda \exp (-i \xi) \equiv \lambda_B=\lambda_F^\ast$. The nomenclature is illustrated in the inset of Fig.~\ref{koch-lattice}(b).  For simplicity, we set $\theta_x$ = $\theta_y$  throughout the calculation. This does not affect the Physics in any way, because all we have to ensure is that, the total Pieerls' phase around a triangle will be $2\pi \Phi/\Phi_0$. Now the values of $\theta_x$ and $\theta_y$ become, $\theta_x = \theta_y = \frac{2 \pi\Phi}{3\Phi_0}$.

On the effectively linear chain (Fig.~\ref{koch-lattice}(c)) one can connect the amplitudes of the wave function on neighboring sites using transfer matrices. In general, corresponds to a difference equation of the form
\begin{equation}
    (E -\epsilon_n) \psi_n = t_{n,n+1} \psi_{n+1} + 
    t_{n,n-1} \psi_{n-1}
    \label{difference1}
\end{equation}
where, $t_{n,n\pm 1}$ imply the hopping amplitudes connecting the $n$-th site to the $n+1$-th and the $n-1$-th sites, one can relate the amplitudes of the wave function $\psi_n$ and $\psi_{n \pm 1}$ through a matrix equation, 
\begin{equation}
\left[ \begin{array}{cc}
\psi_{n+1}\\
\psi_{n}\end{array}\right] = \mathcal{M}_n 
 \left[\begin{array}{cc} \psi_{n}\\
\psi_{n-1}\end{array}\right]
\label{transfer-mat1}
\end{equation}
or, 
\begin{equation}
\left[ \begin{array}{cc}
\psi_{N}\\
\psi_{N-1}\end{array}\right] = \prod_{n=N-1}^1\mathcal{M}_n 
 \left[\begin{array}{cc} \psi_{1}\\
\psi_{0}\end{array}\right]
\label{transfer-mat2}
\end{equation}
for a chain of atoms $N$-sites long. $\psi_1$ and $\psi_0$ provide the desired initial conditions. 
The {\it transfer matrix} $\mathcal{M}_n$ is given by, 
\begin{equation}
    \mathcal{M}_n = \left[ \begin{array}{cccccccccccccccc}
 \frac{E-\epsilon_n}{t_{n,n+1}} & -\frac{t_{n,n-1}}{t_{n,n+1}}\\
 1 & 0\\ 
\end{array} \right ]
\end{equation}
In our case, the hopping amplitudes $t_{n,n \pm 1}$ will have to be appropriately replaced by $\lambda_F$, $\lambda_B$, or simply $t$, while $\epsilon_n$ will be $\epsilon_\beta$, $\epsilon_\delta$, or simply $\epsilon$ for the sites marked `$\beta$', `$\delta$' and `$C$' in Fig.~\ref{koch-lattice}(c). 

Following Fig.~\ref{koch-lattice}(c), we identify three {\it blocks} of sites. Block `1' consists of the pair $\beta\beta$, block `2' consists of a single site of type `C', and block `3' is built out of the quadruplet $\beta\delta\delta\beta$. Three different transfer matrices can be worked out corresponding to these three blocks. They are, 
\begin{eqnarray}
 \mathcal{M}_1 & = & \left[ \begin{array}{cc}
 \frac{(E-\epsilon_\beta)}{t} & -\frac{\lambda_B}{t} \\
1 & 0 \\
\end{array}
\right ] \left[ \begin{array}{cc}
 \frac{(E-\epsilon_\beta)}{\lambda_F} & -\frac{t}{\lambda_F} \\
1 & 0 \\
\end{array}
\right ] \nonumber \\
\mathcal{M}_2 & = & \left[ \begin{array}{cc}
 \frac{(E-\epsilon)}{t} & -1 \\
1 & 0 \\
\end{array}
\right ] \nonumber \\
\mathcal{M}_3 & = & \left[ \begin{array}{cc}
 \frac{(E-\epsilon_\beta)}{t} & -\frac{\lambda_B}{t} \\
1 & 0 \\
\end{array}
\right ] \left[ \begin{array}{cc}
 \frac{(E-\epsilon_\delta)}{\lambda_F} & -\frac{\lambda_B}{\lambda_F} \\
1 & 0 \\
\end{array}
\right ]^2
\left[ \begin{array}{cc}
 \frac{(E-\epsilon_\beta)}{\lambda_F} & -\frac{t}{\lambda_F} \\
1 & 0 \\
\end{array}
\right ]\nonumber\\
\label{matrices}   
\end{eqnarray}
The string of matrices in the product $\prod_n \mathcal{M}_n$ corresponding to Fig.~\ref{koch-lattice}(c) now reads, 
\begin{equation}
    \prod_n \mathcal{M}_n = \mathcal{M}_2 \mathcal{M}_1 \mathcal{M}_2 \mathcal{M}_3 \mathcal{M}_2 \mathcal{M}_1 \mathcal{M}_2
    \label{string}
\end{equation}
Needless to say, if one considers a KF of arbitrarily large generation, the effectively linear chain will appear indefinitely long and consequently, the string of $\mathcal{M}_n$ will consist of an indefinitely long sequence of $\mathcal{M}_1$, $\mathcal{M}_2$ and $\mathcal{M}_3$. The {\it order} of appearance of these matrices will obviously be dictated by the construction of the fractal.

We now make the most important observation. Let us work out the commutators $\left[ \mathcal{M}_2, \mathcal{M}_1\right]$ and $\left[ \mathcal{M}_2, \mathcal{M}_3\right]$. These turn out to be, 

\begin{equation}
\left[ \mathcal{M}_2, \mathcal{M}_{1(3)}\right] = \left[ \begin{array}{cccccccccccccccc}
 0 & \Gamma_{2,1(3)}\\
\Gamma_{2,1(3)} & 0\\
\end{array}
\right ] 
\label{2-1-3}
\end{equation}

\begin{figure*}[ht]
\centering
(a)\includegraphics[width=0.6\columnwidth]{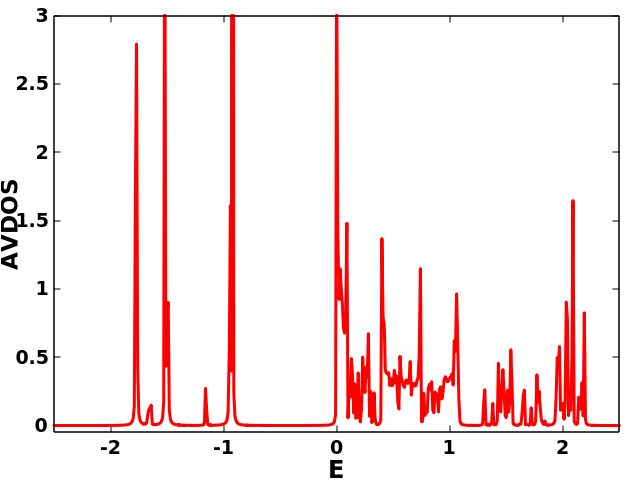}
(b)\includegraphics[width=0.6\columnwidth]{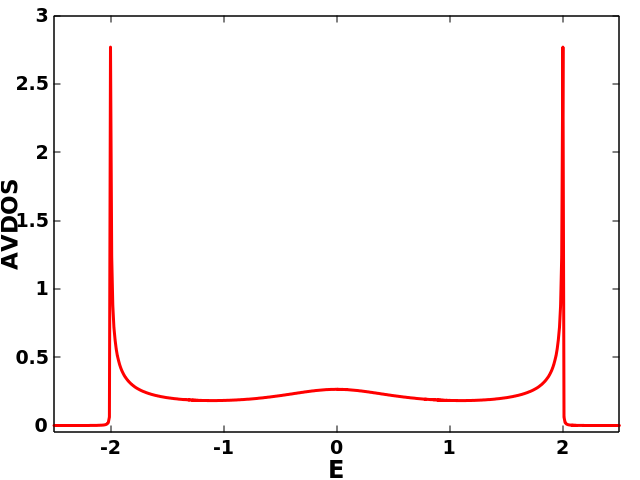}
(c)\includegraphics[width=0.6\columnwidth]{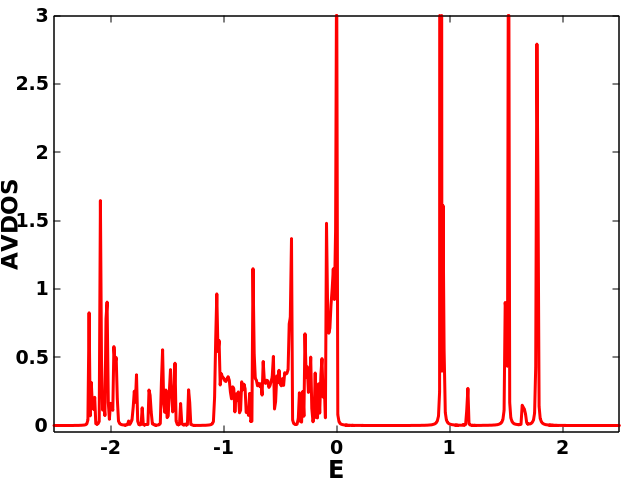}
(d)\includegraphics[width=0.6\columnwidth]{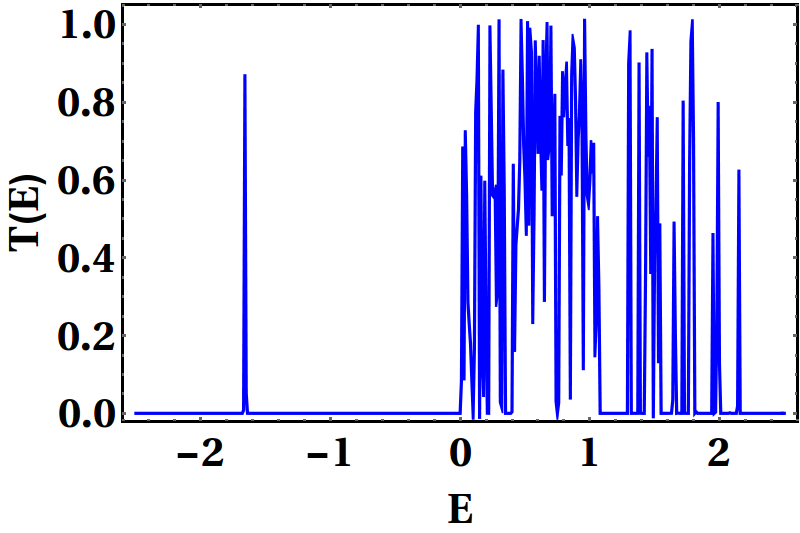}
(e)\includegraphics[width=0.6\columnwidth]{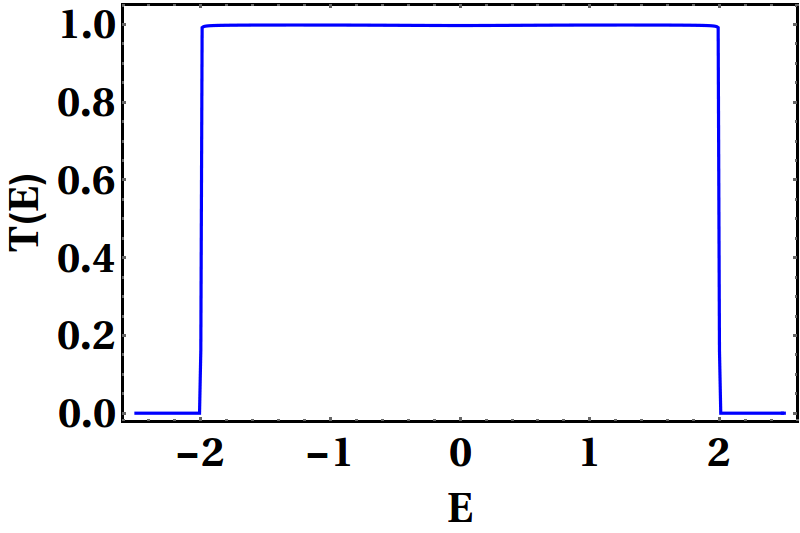}
(f)\includegraphics[width=0.6\columnwidth]{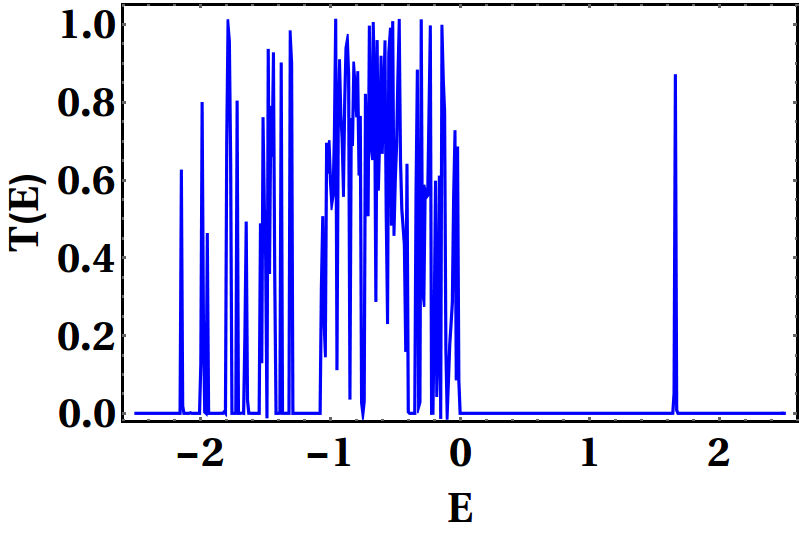}
(g)\includegraphics[width=0.6\columnwidth]{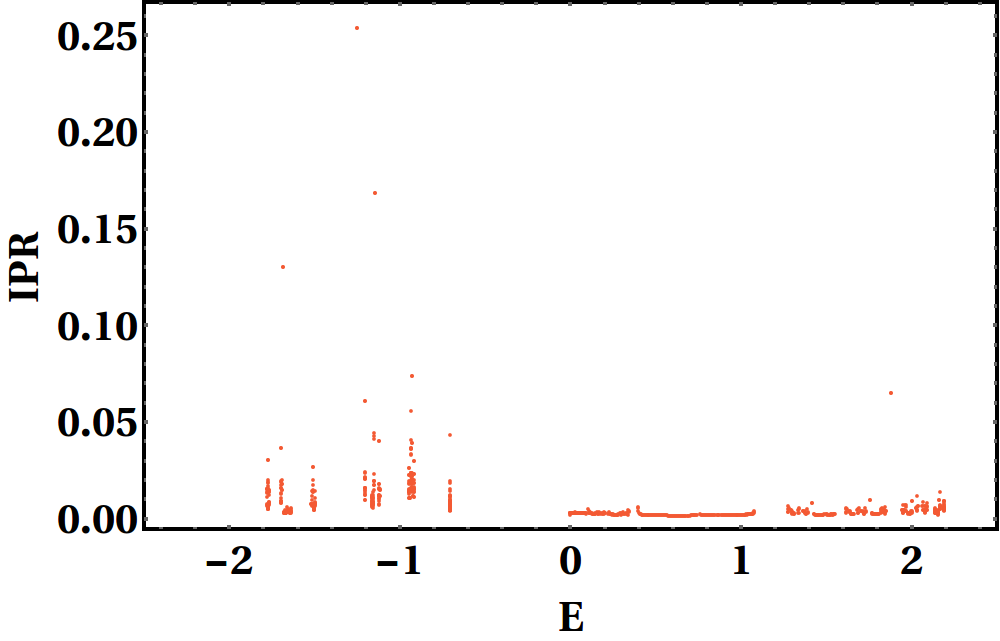}
(h)\includegraphics[width=0.62\columnwidth]{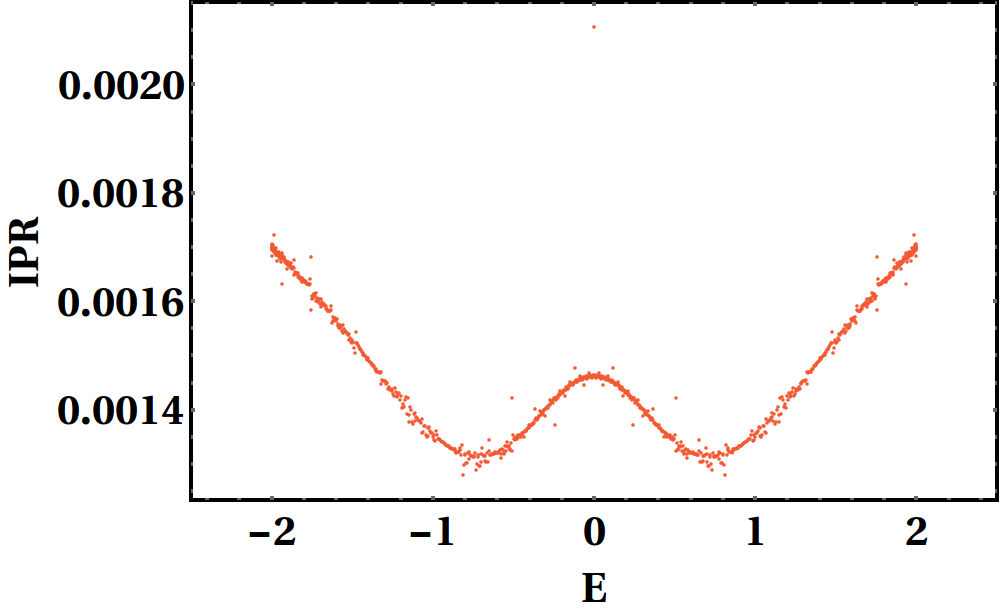}
(i)\includegraphics[width=0.6\columnwidth]{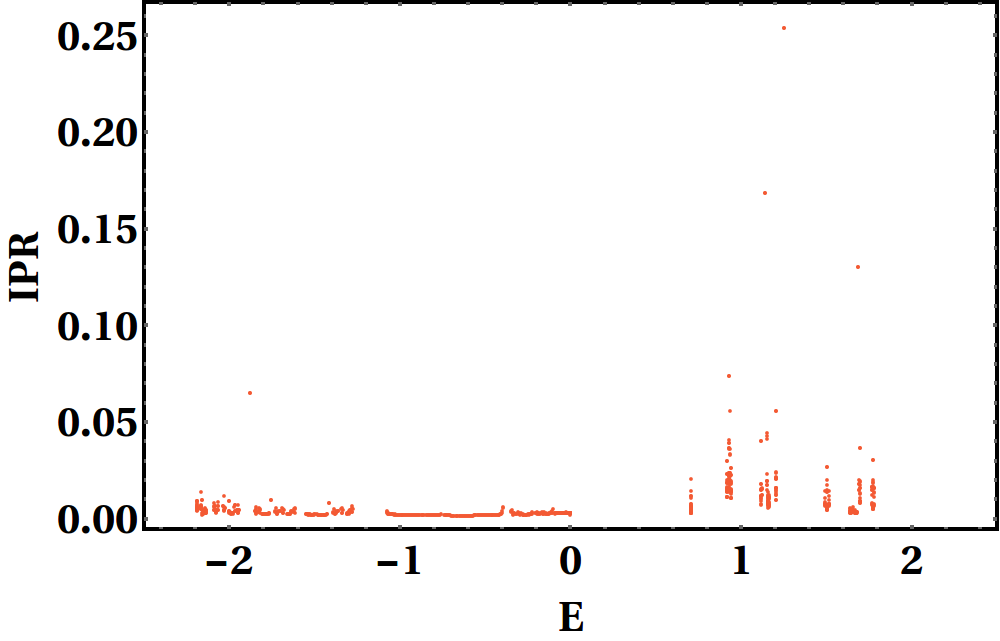}
\caption{(Color online) (a) - (c) display the average density of states of an infinite koch fractal, obtained using the decimation renormalization method outlined in appendix A. (d) - (f)  depict the transmission co-efficient plotted against the energy, and (g) - (i) show the variation of the inverse participation ratio (IPR) with energy. The transmission coefficient and the IPR have been calculated for a Koch fractal in its 5th generation. The parameters, for all the figures, are chosen as, $ \epsilon = 0$, $t = 1$, $x = y = \frac{1}{\sqrt{2}}$. The magnetic flux $\Phi = 0$ in (a),(d),(g), $\Phi = \frac{1}{4} \Phi_0$ in (b), (e), (h), and $\Phi = \frac{1}{2} \Phi_0$ in (c), (f), (i). $\Phi_0$ has been set equal to unity.}  
\label{dos}
\end{figure*}

where, 
\begin{eqnarray}
\Gamma_{2,1} & = & -\frac{e^{i\frac{4 \pi\Phi}{3\Phi_0}} \left [ 2 x^2y \cos{\frac{2 \pi\Phi}{\Phi_0}}+(E-\epsilon)(-t^2+x^2+y^2) \right ]}{t \left [x^2+e^{i\frac{2 \pi\Phi}{\Phi_0}}(E-\epsilon)y \right]}\nonumber\\
\Gamma_{2,3} & = & -\frac{e^{i\frac{4 \pi\Phi}{\Phi_0}}\left [(E-\epsilon)(-t^2+x^2+y^2)+2x^2y \cos{\frac{2 \pi\Phi}{\Phi_0}}\right]\xi}{t \left [x^2+e^{i\frac{2 \pi\Phi}{\Phi_0}}(E-\epsilon)y \right]^3}\nonumber\\
 \label{com-ele}
\end{eqnarray}
With, $ \xi = E^4-4E^3\epsilon+6E^2\epsilon^2-4E\epsilon^3+\epsilon^4-4E^2x^2+8E\epsilon x^2-4\epsilon^2x^2+3x^4-(E-\epsilon)^2y^2-2(E-\epsilon)x^2y\cos{\frac{2 \pi\Phi}{\Phi_0}}$.
From Eqs.~\eqref{2-1-3} and Eqs.~\eqref{com-ele}, it is clear that a selection of $x^2+y^2=t^2$ and $\Phi=\frac{1}{4}\Phi_0$ makes $\left[ \mathcal{M}_2, \mathcal{M}_{1(3)}\right] = 0 $   {\it independent of energy} $E$. This implies that, in the effectively linear chain in Fig.~\ref{koch-lattice}(c) and its longer versions (of any length actually), we can rearrange the clusters $1$, $2$, or $3$ in {\it any} sequence. For example, we can switch their positions to form a {\it completely periodic} pattern.  The corresponding string of the transfer matrices are then naturally rearranged, for example, as $\prod_n \mathcal{M}_n = [\mathcal{M}_1]^2 [\mathcal{M}_2]^4 [\mathcal{M}_3]$ corresponding to Fig.~\ref{koch-lattice}(c). On a linear chain, derived out of an {\it infinitely large} Koch fractal geometry, we then have an infinitely long string of matrices representing the $\beta\beta$ cluster, the isolated $C$ site, and the $\beta\delta\delta\beta$ cluster. The commutation allows us to rearrange the corresponding matrix product as, 
\begin{equation}
    \prod_{n=-\infty}^{\infty} \mathcal{M}_n = 
   \left ( \prod_{i=-\infty}^{\infty} \mathcal{M}_1^i \right ) \cdot \left (\prod_{j=-\infty}^{\infty} \mathcal{M}_2^j \right ) \cdot 
    \left (\prod_{k=-\infty}^{\infty} \mathcal{M}_3^k \right )
    \label{newstring}
\end{equation}
In Eq.~\eqref{newstring} above, each of the `sub-strings' of  $\mathcal{M}_1$, $\mathcal{M}_2$ or $\mathcal{M}_3$ has an infinite number of the basic matrices, which account for the subsets of the $\beta\beta$ clusters, the $C$-sites and the $\beta\delta\delta\beta$ clusters of the total infinite chain. The concentrations of each cluster (or, equivalently, the kinds of sites) will be used in the next subsection while evaluating the average density of states.

The string of matrices on the right hand side of Eq.~\eqref{newstring} now represents a lattice geometry in which one has a {\it completely periodic} array of $\beta\beta$ clusters (The $\mathcal{M}_1$ string), followed by an infinite string of $C$-sites ($\mathcal{M}_2$ string) and then the $\beta\delta\delta\beta$ clusters of atomic sites ($\mathcal{M}_3$ clusters). For a KF lattice of infinite size, each such individual chain of clusters is infinitely long. The commutation $[\mathcal{M}_i, \mathcal{M}_j] =0$ for $i,j=1,2,3$. The commutator bracket vanishes independent of the energy $E$ of the incoming electron. As a consequence, the individual blocks $\beta\beta$, $C$ and $\beta\delta\delta\beta$ can be distributed in a completely periodic manner for {\it all} energy eigenvalues. Hence we expect an {\it absolutely continuous} spectrum with {\it all states extended}, Bloch-like.

In this context, it should be appreciated that the existence of the extended eigenstates, as discussed above is a non-trivial addition to some of the earlier fundamental works on tight binding formalism of hierarchical lattices in one dimension~\cite{roman,ceccato} where the Cantor set spectrum of the Hamiltonian and the quantum states of the propagating excitation had been studied in great details and the multifractal character of the wave functions was explained. In another illuminating study of a hierarchical potential model~\cite{kunz} in one dimension and beyond, a Cantor set energy spectrum with singular continuous character was discussed and the appearance of a continuous distribution of eigenvalues for a certain range of the energy spectrum was discussed when the hierarchy parameter was chosen to have a special range of values. However, a complete turnover of the spectrum from a cantor-set character to an absolutely continuous one, populated by extended eigenstates only, is something new to our understanding, and was not observed in a hierarchical KF before, to the best of our knowledge.

\subsection{The density of states}
The eigenvalue spectrum is conveniently captured in the average density of states (ADOS) of the system under study. Here, exploiting the self-similarity of the KF structure we have adopted a real space renormalization group (RSRG) method~\cite{southern} to obtain the average density of states of 
an infinite KF. The ADOS is given by, 
\begin{equation}
    \rho (E) = w_A \rho_A(E) + w_B \rho_B(E) + w_C \rho_C(E) 
    + w_D \rho_D(E)
    \label{ados}
\end{equation}
where, $\rho_J$ ($j=A,B,C,D$) is the `local' density of states at the site of type $j$, and the respective concentrations $w_j$ ($j=A,B,C,D$) are obtained as, $w_A=w_B=1/3$, and $w_C=w_D=1/6$. The details of the RSRG recursion relations to obtain the local Green's functions and hence the density of states is outlined in Appendix $A$.

In Fig.~\ref{dos}(a) - (c) we show the average density of states (AVDOS) of an infinite Koch fractal, obtained by the RSRG decimation scheme elaborated in Appendix $A$. Three prototype AVDOS profiles are shown. In Fig.~\ref{dos}(a) and (c) the AVDOS is shown when the flux threading each triangular plaquette are $\Phi=0$ and $\Phi=(1/2) \Phi_0$ respectively. The spectrum in each case shows a rugged landscape for the AVDOS, typical of such a fractal lattice~\cite{macia,arunava-koch1,arunava-koch2}. It is noticed that there are some patches where the AVDOS appears to be continuous, and the corresponding end-to-end transmission coefficient (see Section IV), shown in panels (d) and (f) also appears to be high, close to unity. This is not unexpected, as the Koch fractal is known to possess (at zero and non-zero flux) {\it extended} but isolated eigenstates at some {\it special}  energy values. Such states, for an infinitely large fractal, can even be infinite in number, as discussed before~\cite{arunava-koch1,arunava-koch2}. 

The remarkable change in the AVDOS profile, brought out under the commutation condition $x^2 + y^2 =t^2$, and $\Phi = \Phi_0/4$, is explicitly seen in Fig.~\ref{dos}(b).  
It can be seen that, the fragmented AVDOS profiles in (a) or (c) now collapse into a single, {\it absolutely continuous} spread of energy eigenvalues $E$ ranging between $[-2,2]$. We have examined numerous combinations of $x$ and $y$, maintaining the condition $x^2 + y^2 =t^2$, and keeping the flux fixed at $\Phi=\Phi_0/4$. In each case, the `extent' of the band is between $[-2,2]$ with $t$ set equal to unity. In fact, the local densities of states (LDOS) at all kinds of sites show the same bandwidth, though the fine structure of the LDOS profile are usually different.

\begin{figure}
    \centering
    \includegraphics[width=.9\columnwidth]{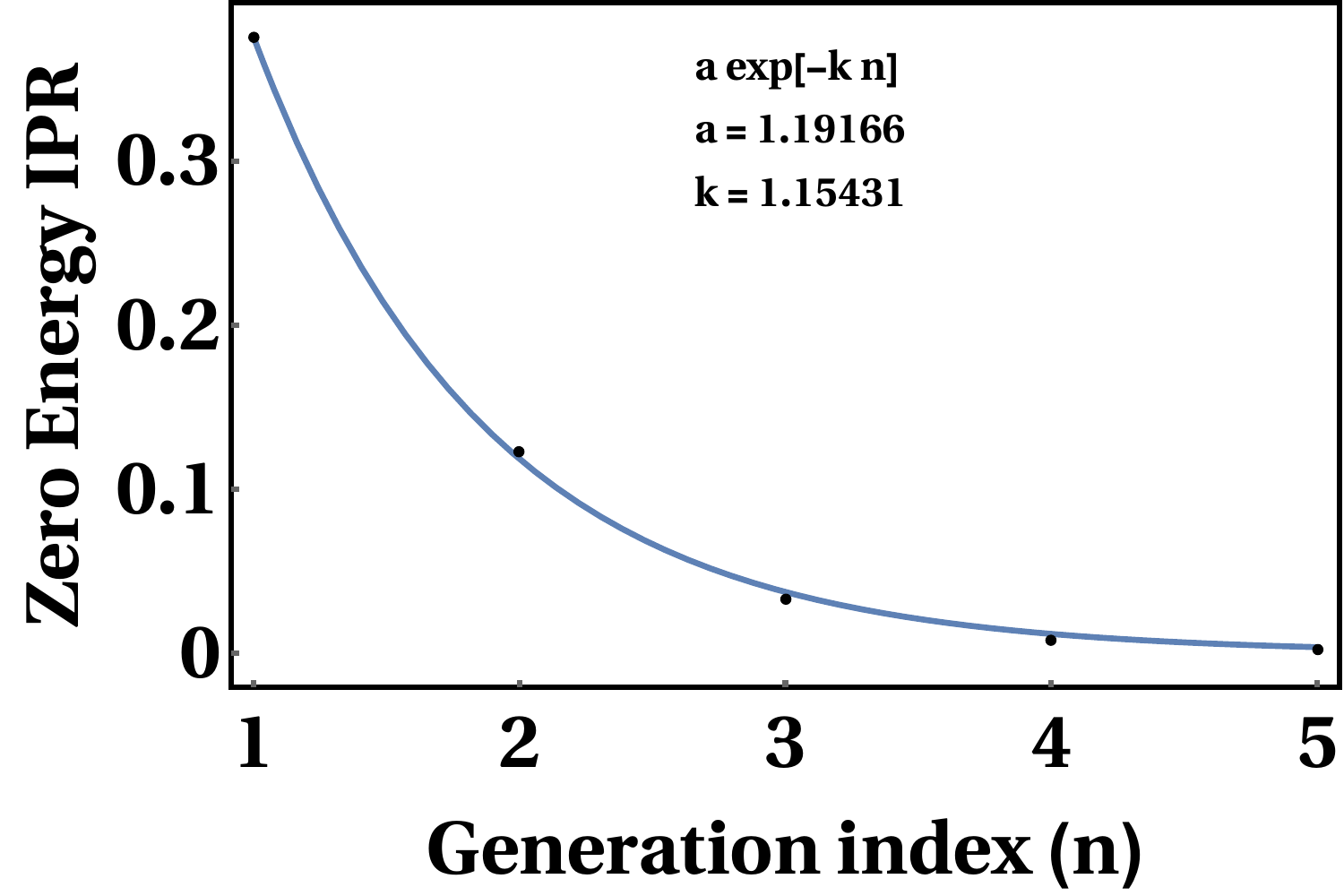}
    \caption{(color online) Variation of zero energy inverse participation ratio (IPR ) with the generation index for a Koch fractal network.  The parameters are chosen as $ \epsilon = 0 $, $ t = 1 $, $ x = \frac{1}{\sqrt{2}}$,  $ y = \frac{1}{\sqrt{2}}$, $\Phi = \frac{1}{4}\Phi_0$.}
    \label{iprmax}
\end{figure}

The absolutely continuous nature of the spectrum corresponds to a set of completely {\it extended} eigenfunctions, a fact that is easily confirmed from the flow of the hopping integrals under the RSRG iterations~\cite{southern} (Appendix). For any energy eigenvalue, picked up at random from anywhere within the energy spectrum in Fig.~\ref{dos}(b), and for {\it any combination} of the set $(x,y)$, satisfying $x^2 + y^2 =t^2$, and with $\Phi=\Phi_0/4$, the numerical values of hopping integrals keep on oscillating, without showing any sign of convergence to zero. This implies that, for any energy within the band, there is a non-zero overlap between the wavefunctions at {\it all} scales of length. This is a confirmatory test for the extendedness of the wave function, a fact that is substantiated by the perfect two-terminal transport, as depicted in Fig.~\ref{dos}(e).

To strengthen the argument regarding the extended character of the eigenstates under the special condition referred to above, we have evaluated the inverse participation ratio (IPR) for increasingly larger finite-sized Koch fractal lattices. IPR is widely used in studying the localization properties of disordered lattices, and we outline the basic formula in Appendix C for the benefit of the reader. 

In Fig.~\ref{dos} (g) - (i) one such case is shown, for three different flux values. The panel (g) for $\Phi=0$ and panel (i) for $\Phi=\Phi_0/2$, finite values of the IPR indicate the existence of localized eigenstates. There are of course regions in the energy axis over which the IPR assumes quite low values, indicating larger localization lengths or even extendedness of the states. This is not surprising as a Koch fractal of the type considered here is already known to possess isolated but closely packed extended eigenstates at discrete values of the energy E~\cite{macia, arunava-koch1,arunava-koch2}. However, as one enforces the resonance (commutation) condition, the gross IPR profile distinctly comes down arbitrarily close to zero, implying that a long enough Koch geometry, with $x^2+y^2=t^2$ and $\Phi=\Phi_0/4$, will support {\it only} extended wavefunctions. We have cross-checked this surmise (for finite systems) by plotting the maximum value of the IPR, which occurs at $E=0$, against the generation number, as shown in Fig.~\ref{iprmax}. The curve clearly shows an exponential decay in the maximum value of the IPR as the system size is increased, with an exponent equal to 1.15431.

\subsection{A unique band at resonance}
\begin{figure}[ht]
\centering
\includegraphics[width=\columnwidth]{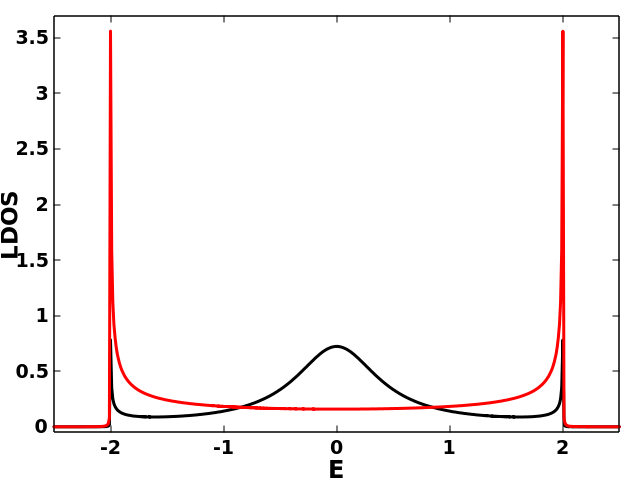}
\caption{(Color online) The local densities of states $\rho_A$ (black line), and $\rho_B$, $\rho_C$ and $\rho_D$ (all merge in the single red line) in a Koch fractal. We have set $x=0.6$, $y=0.8$ (satisfying $x^2+y^2=t^2$. The flux trapped is set equal to $\Phi_0/4$. The on-site potential is $\epsilon=0$ everywhere, and $t$ has been set equal to unity. Energy is measured in units of $t$.}  
\label{band}
\end{figure}

As the conditions $x^2+y^2=t^2$ and $\Phi=\Phi_0/4$ are enforced, the spectrum of eigenvalues is confined between $E=\epsilon-2t$ and $\epsilon+2t$ and does not spill over this range. In Fig.~\ref{band} we show the local density of states (LDOS) at the four varieties of sites $A$, $B$, $C$, and $D$. We have intentionally set $x=0.6$ and $y=0.8$ just as a variation from the values used earlier. The flux is ket at $\Phi=\Phi_0/4$. The red curve, representing $\rho_C$, exactly resembles the LDOS of a perfectly periodic chain of identical atomic sites with on-site potential $\epsilon=0$ and nearest neighbor hopping integral $t=1$. Interestingly, the LDOS at two other sites $B$ and $D$, viz, $\rho_B$ and $\rho_D$ exactly match the profile of $\rho_C$ both qualitatively and quantitatively, and hence merge completely with $\rho_C$. In Appendix B we give an analytical proof that this is indeed the case. The black curve shows the variation of the LDOS at the site of type $A$, that is, $\rho_A$. The band-width shown by $\rho_A$ matches those shown by $\rho_B$ or $\rho_C$ or $\rho_D$ as seen in Fig.~\ref{band}. The bulk of the $\rho_A$ differs from the others, but the spectrum of $\rho_A$ is absolutely continuous.

\section{Transmission Characteristics}

\begin{figure}[ht]
\centering
\includegraphics[width=\columnwidth]{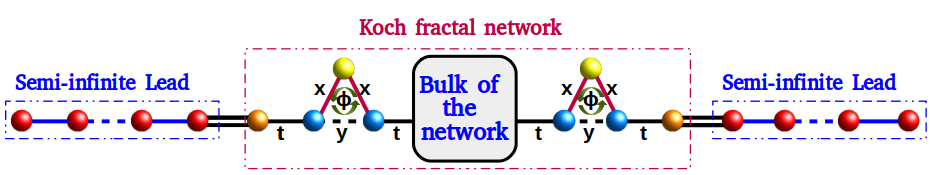}
\caption{(Color online) Schematic diagram of the system under study coupled to the perfectly ordered semi-infinite leads at the two ends.}  
\label{lead}
\end{figure}

We have calculated the two-terminal transmission coefficient of finite Koch fractals, and present the results here for a fifth-generation fractal having a total of $1025$ atomic sites. The `sample' is clamped between two semi-infinite periodic `leads' (red atoms), and is schematically shown in Fig.~\ref{lead}. 
We adopt the standard Green’s function technique. The Green's function of the whole system is defined as $ G = (E - H )^{-1}$. The full Hamiltonian $H$ of the combined system consisting of the finite-sized fractal network and semi-infinite leads is symbolically expressed as, 
\begin{equation}
    H  =  H_S + H_{1}+H_{2}+H_{S1}+H_{S2}+H^\dagger_{S1}+H^\dagger_{S2}
\end{equation}
Where $H_S$ is the tight binding Hamiltonian of the fractal system.

The Hamiltonian of the semi-infinite lead systems is described by $H_1(H_2)$. The terms $H_{S1}(H^\dagger_{S1})$ and $H_{S2}(H^\dagger_{S2})$ account for the coupling between the first and second leads and the fractal system respectively. The effective Green’s function of the fractal system can be defined as~\cite{fisher,supriya},

\begin{equation}
    G = ( E - H_S - \Sigma_1 - \Sigma_2)^{-1}
\end{equation}
where, $\Sigma_{1(2)} = H_{S1(2)}G_{1(2)}H^\dagger_{S1(2)} $ are the self-energy correction terms that arise due to the attachment of the leads to the system.  $G_{1(2)} = [E - H_{1(2)}]^{-1}$ corresponds to the Green’s functions of the two leads. Once the self-energy is calculated, it is straightforward to obtain the coupling function, $\Gamma_{1(2)}(E)$, viz.

\begin{equation}
    \Gamma_{1(2)}(E) = i \left[ \Sigma^{ret}_{1(2)}(E) - \Sigma^{adv}_{1(2)}(E)  \right]
    \label{coupling}
\end{equation}
where, the advanced self-energy $\Sigma^{adv}_{1(2)}(E)$ and  the retarded self-energy $\Sigma^{ret}_{1(2)}(E)$ are Hermitian conjugates of each other. Eq.~\eqref{coupling} therefore, can then be written as, 

\begin{equation}
    \Gamma_{1(2)}(E) = 2 Im \left[ \Sigma^{ret}_{1(2)}(E)  \right]
    \label{coupling-1}
\end{equation}

The  expression for the transmission coefficient between the first and second lead, as a function of the system-lead coupling, can be written as~\cite{supriya}, 

\begin{equation}
    T(E) = Tr \left[ \Gamma_1(E) G^{ret} \Gamma_2(E) G^{adv}  \right]
    \label{trans}
\end{equation}

The transmission coefficient as a function of energy for a Koch fractal of 1025 sites is displayed in panels (d), (e), and (f) in Fig.~\ref{dos}. Away from the resonance condition, $T(E)$ exhibits a rugged landscape, occasionally hitting the value unity (or close to unity), and remaining low or even becoming zero at energies at which either one encounters localized states or no states at all. The central panel (e) corresponds to the resonance condition, viz, $x^2+y^2=t^2$ and $\Phi=\Phi_0/4$, and shows that the entire fractal is completely transparent to an incoming electron with {\it any energy} within $E=-2$ and $E=2$. 

To evaluate the transmission coefficient numerically, one has to take a reasonable size of the KF lattice (the 1025 site long chain taken here seems enough for the purpose) and set the imaginary part of energy $E$ while evaluating the Green's functions~\cite{supriya} to a low value (the imaginary part in energy is needed to lift the poles of the Green's function off the real axis, and should tend to zero). For numerics, we choose the imaginary part in energy as $10^{-6}$. The result is in exact correspondence with the absolutely continuous AVDOS shown in panel (b) in the same figure. In fact, all the transmission spectra shown, are in exact correspondence with the AVDOS diagrams, as they should be. 

The perfect transmission under the resonance condition is not unexpected. For example, in the extreme case of $y=0$, the resonance condition boils down to $x=t$, and one has a perfectly periodic one-dimensional lattice for which the density of states is the red curve in Fig.~\ref{band}. The transmission should be perfect. Addition of $y$, and imposing the resonance condition changes the profile of the AVDOS, but the absolute continuity of the spectrum is not affected. Therefore, we still expect a perfect transmission, and that is precisely what we see here.

\section{A subtle universality}
The commutation of the transfer matrices corresponding to the three basic building blocks of the Koch fractal shown in Fig.~\ref{koch-lattice} gives rise to a few interesting observations.

First, let us point out that, essentially the geometry considered here is a linear one. That is, one can stretch the Koch fractal and get a line punctuated with single triangles (block 1), isolated `dots' (block 2), and the three-triangle cluster (block 3). The condition $x^2 + y^2 =t^2$ implies that one can play around with an infinite variety of combinations $(x,y)$ which lie on the perimeter of a circle centered at $(x=0, y=0)$, and having a radius equal to $t$. For a given value of $t$, all such different $(x,y)$ combinations actually refer to Koch fractals with different {\it microscopic details}. Even, within a single Koch fractal geometry one can assign different combinations of $x$, $y$, and $t$, yet maintaining the sacred equality $x^2 + y^2 =t^2$, and setting $\Phi=\Phi_0/4$. The eigenvalue spectra and transmission properties will definitely look different. Yet, the implementation of the resonance condition makes the local density of states at the $B$ or the $D$ sites completely identical, and both resemble the LDOS of a perfectly periodic array of atoms, with a constant on-site potential $\epsilon$ and a constant nearest neighbor hopping integral $t$. The transmission spectrum at the resonance will be the {\it same} for {\it all} such lattices. Making $x$ and $y$ different from $t$, in a sense, defines a kind of {\it local distortion} in the lattice. Hence, the above arguments hint to the existence of a kind of {\it universality} among microscopically different Koch fractal lattice models of such kind.\\
Second, it is to be appreciated that, once the commutation is achieved at the onset of the resonance condition, the arrangement of blocks 1, 2, and 3 becomes completely irrelevant, and one can even arrange these in a completely disordered fashion. However, the qualitative or the quantitative characteristics of the LDOS, as well as the bandwidth will not change at all. The quantities $\rho_B$ or $\rho_D$ will look identical as long as $x^2+y^2 = t^2$, with $\Phi=\Phi_0/4$.\\
These observations indeed call for a deeper analysis in regard of a possible universality class among such decorated lattices, a task that will be undertaken later.

\section{Concluding Remarks}
In this work, we have analytically worked out the criteria under which a flux-threaded Koch fractal lattice becomes completely transparent to an incoming electron having {\it any energy} lying within the eigenvalue spectrum. The results bring out a non-trivial variation over the conventional cases of disorder-driven localization phenomena. The results differ completely from the previous pioneering works related to the random dimer model or its variants in randomly disordered or quasiperiodically ordered lattices. Studies of the inverse participation ratio and the two terminal transport corroborate the first results obtained from transfer matrix calculations in every respect. We also point out that any variation in the system parameters, especially the hopping amplitudes around local atomic plaquettes, that are consistent with the {\it resonance} condition, leads to the same bandwidth and local densities of states at a subset of sites. This opens up a new kind of universality class for an entire group of lattices, including geometrically randomised arrangements, comprising similar triangular plaquettes and dots. 

\section{Acknowledgments}
SB is thankful to the Government of West Bengal for the SVMCM Scholarship (WBP221657867058). 

\appendix
\section{}
We explain here the RSRG scheme used to determine the average density of states of an infinite Koch fractal lattice.
\begin{figure}[ht]
\centering
\includegraphics[width=.9\columnwidth]{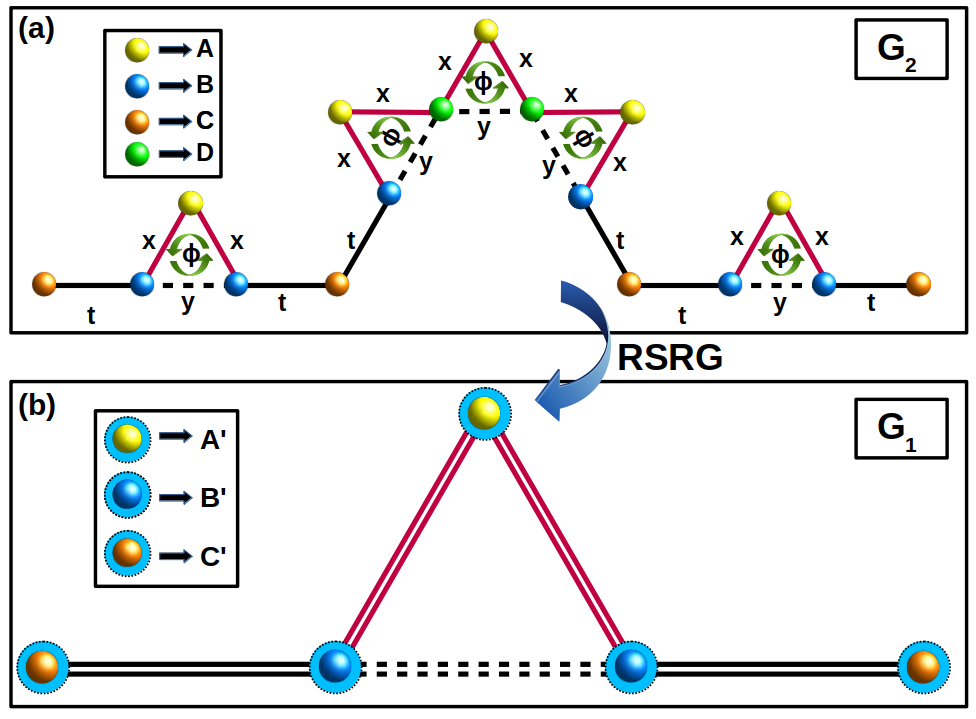}
\caption{(Color online) Decimation renormalization of a portion of an infinitely large Koch fractal. The undecimated sites are encircled, and they have the renormalized on-site potentials, as explained in the text. The energy dependent nearest neighbor hopping integrals are along the double bonds in the figure. They are $x_{F(B)}'$, $y_{F(B)}'$ and $t_{F(B)}'$, and are shown by the red, black dotted, and the black colored double lines respectively.} 
\label{rsrg}
\end{figure}
Let us refer to Fig.~\ref{rsrg} that explains the scaling of a KF in a certain generation (Fig.~\ref{rsrg}(a)) to its previous generation. The nomenclature of the sites and the hopping integrals remain the same as that in the main text. A subset of the vertices $A$, $B$, $C$, and $D$ are decimated using the difference equations Eq.~\eqref{difference} and
Eq.~\eqref{difference1}. This results in the renormalized version of the lattice, as shown in Fig.~\ref{rsrg}(b), where the renormalized on-site potentials and the hopping integrals are given by, 


    \begin{eqnarray}
    \epsilon_A' & = & \epsilon_A + \left [ \frac{x_F ( p_2^\ast + p_3^\ast p_2) + x_B ( p_2 + p_2^\ast p_3)}{1-p_3^\ast p_3} \right ] \nonumber \\
    \epsilon_B' & = & \epsilon_C + \frac{t_F t_B (E-\epsilon_1)}{\delta_1} + \frac{t_F t_B}{E - \epsilon_1-q_1 \tau_F} \nonumber \\
    \epsilon_C' & = & \epsilon_C + \frac{2 t_F t_B (E-\epsilon_1)}{\delta_1} \nonumber \\
    \epsilon_D' & = & \epsilon_C + \frac{t_F t_B}{E - \epsilon_1 -q_1^\ast \tau_B} + \frac{t_F t_B}{E - \epsilon_1 - q_1 \tau_F} \nonumber \\
    x_F' & = & \frac{q_2^\ast t_B \tau_B}{E - \epsilon_1 -q_1^\ast \tau_B} \nonumber \\
    y_F' & = & \frac{q_3 t_F \tau_F}{E - \epsilon_1 -q_1 \tau_F} \nonumber \\
    t_F' & = & \frac{t_F^2 \tau_F}{\delta_1}
    \label{recursion}
\end{eqnarray}
The symbols used in the recursion relations given in Eq.~\eqref{recursion}, when written elaborately, read

\begin{eqnarray}
    \epsilon_1 & = & \epsilon_B + \frac{x_F x_B}{E - \epsilon_A} \nonumber \\
    \epsilon_2 & = & \epsilon_D + \frac{x_F x_B}{E - \epsilon_A} \nonumber \\
    \tau_F & = & y_F + \frac{x_B^2}{E-\epsilon_A} \nonumber \\
    \tau_B & = & y_B + \frac{x_F^2}{E-\epsilon_A} \nonumber \\
    \delta_1 & = & (E - \epsilon_1)^2 - \tau_F \tau_B \nonumber \\
    \delta_2 & = & (E-\epsilon_1) (E-\epsilon_2) - \tau_F \tau_B \nonumber \\
    p_1 & = & \frac{t_F \tau_F}{\delta_2} \nonumber \\
    p_2 & = & \frac{(E-\epsilon_1) x_F}{\delta_2} \nonumber \\
    p_3 & = & \frac{(E-\epsilon_1) y_B}{\delta_2} \nonumber \\
    q_1 & = & \frac{\tau_B}{E-\epsilon_2-p_3 y_F} \nonumber \\
    q_2 & = & \frac{x_B + p_2 y_F}{E-\epsilon_2-p_3 y_F} \nonumber \\
    q_3 & = & \frac{p_1 y_F}{E - \epsilon_2-p_3 y_F} 
    \label{details}
\end{eqnarray}

The complex conjugates of the above quantities are easily obtained.
It should be appreciated that though the hopping integral $t$ along the backbone is real initially, on renormalization it picks up a phase and hence we have to deal with its complex versions, viz, the `forward' (F) and the `backward' (B). The initial values of the other hopping integrals are, $x_F = x \exp (i\theta_x)$, and $y_F=y \exp (i\theta_y)$. The complex conjugate $x_B$ and $y_B$ are easily obtained.

The set of Eqs.~\eqref{recursion} is iterated with a small imaginary part $\eta$ added to the energy $E$, until the effective nearest neighbor hopping integrals vanish, and one is left with the {\it fixed point} values of the on-site potentials, viz, $\tilde \epsilon_j$, with $j=A, B, C$, and $D$. The {\it local} Green's function at any vertex is then obtained from the standard expression $G_{jj} = (E-\tilde{\epsilon_j})^{-1}$, and the local density of states at the $j$-th vertex is obtained as, 
\begin{equation}
    \rho_j =Lim_{\eta \rightarrow 0} \left [-\frac{1}{\pi} Im G_{jj}(E + i \eta) \right ]
    \end{equation}

\section{}
\begin{figure}[ht]
\centering
\includegraphics[width=\columnwidth]{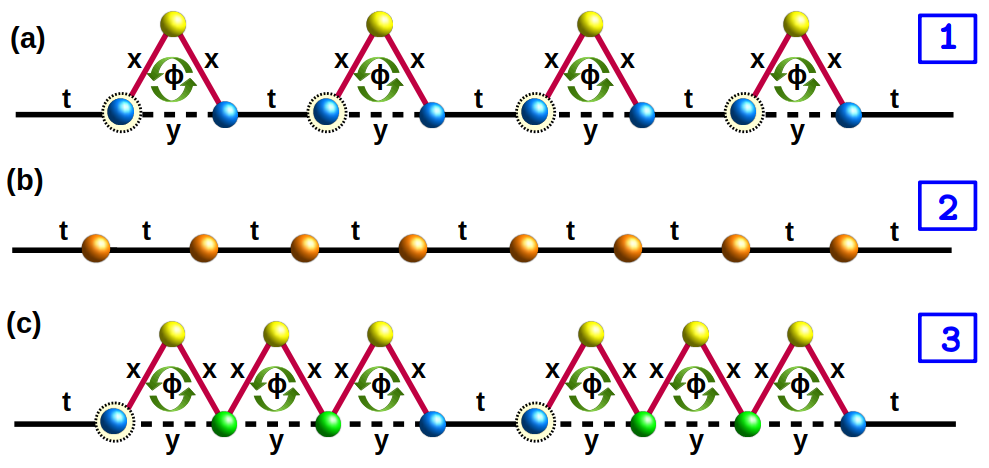}
\caption{(Color online) Ordered equivalents of the Koch fractal geometry when the commutation conditions are imposed. The three basic building blocks, namely, ``block 1", ``block 2" and ``block 3" in Fig.~\ref{koch-lattice} (c) are shown to form three different infinitely periodic lattices themselves. The counterclockwise arrows in the triangles show the convention chosen to define the `forward' or the `backward' hopping, as before. The fine structure of ``block 3" is shown in (c).}  
\label{decoupled}
\end{figure}

As the transfer matrices commute, independent of energy $E$, when we set the resonance condition, each sub-string of the transfer matrices given in the parentheses of Eq.~\eqref{newstring} represents a perfectly periodic arrangement of the blocks $1$, $2$ and $3$ respectively. Each such arrangement is actually an infinitely large ordered one-dimensional {\it crystal}, with designer unit cells, as shown in Fig.~\ref{decoupled} (a), (b), and (c). Therefore, under the resonance condition, one needs a careful analysis of the LDOS or the AVDOS of each such chain to understand the global band of the parent Koch fractal.

For example, Fig.~\ref{decoupled} (b) is just an ordinary infinite array of identical atomic sites with constant on-site potential $\epsilon$ and the nearest neighbor hopping integral $t$. The energy bands extend from $E=\epsilon-2t$ to $E=\epsilon+2t$, and all the eigenfunctions are extended, Bloch functions. The local (or average) density of states for this chain can be analytically obtained, and is given by,
\begin{equation}
\rho_C = 
\frac{1}{\pi} \frac{1}{\sqrt{4t^2-(E-\epsilon)^2}}
\end{equation}
The van-Hove singularity is obvious at the band-edges, viz, at $E = \epsilon \pm 2t$.

The remaining arrays depicted in Figs.~\ref{decoupled}(a) and (c) can easily be renormalized into effective chains of identical atomic sites, by decimating out a subset of the vertices of the triangles. The $A$ sites (yellow) and half of the $B$ sites (blue) in Fig.~\ref{decoupled}(a) are decimated. The `surviving' $B$ sites are encircled in both figures, and they form an effectively one-dimensional periodic chain. Both the on-site potential $\epsilon_1$ and the nearest neighbor hopping integral $t_1$, connecting the neighboring encircled sites on this effective one-dimensional array of renormalized $B$-sites, are energy dependent. $t_1$ is complex because of the flux trapped in the triangle. A similar decimation process executed on Fig.~\ref{decoupled}(c) results in an effectively linear lattice of a subset of the $B$ sites, which now have renormalized, energy-dependent on-site potentials $\epsilon_3$ and complex nearest neighbor hopping integral $t_3$. 

The densities of states at the encircled sites in both Fig.~\ref{decoupled}(a) and (c) can now be obtained analytically. An easy but cumbersome algebra leads to the following results:

\begin{equation}
\rho_{1(3)}(E) = \frac{1}{\pi} \frac{1}{\sqrt{4{t_{1(3)}}^2-(E-\epsilon_{1(3)})^2}}
\label{dosac}
\end{equation}
where, 
\begin{eqnarray} 
\epsilon_{1} & = & \epsilon_\beta +\frac{t^2+\lambda^2}{E- \epsilon_\beta} \nonumber\\
t_{1}^2 & = & \frac{t^2 \lambda^2}{(E- \epsilon_\beta)^2}\nonumber\\
\epsilon_{3} & = & \epsilon_\beta+\frac{\lambda^2(E-\epsilon_\delta)}{(E-\epsilon_\delta)^2-\lambda^2} + \frac{t^2+\frac{\lambda^6}{[(E-\epsilon_\delta)^2-\lambda^2]^2}}{E-\epsilon_\beta-\frac{\lambda^2(E-\epsilon_\delta)}{(E-\epsilon_\delta)^2-\lambda^2}}\nonumber\\
t_{3}^2 & = & \frac{t^2\frac{\lambda^6}{[(E-\epsilon_\delta)^2-\lambda^2]^2}}{[E-\epsilon_\beta-\frac{\lambda^2(E-\epsilon_\delta)}{(E-\epsilon_\delta)^2-\lambda^2}]^2} \nonumber\\
\lambda^2 & = & \frac{y^2(E-\epsilon)^2+x^4+2x^2y(E-\epsilon) \cos{\frac{2\pi\Phi}{\Phi_0}}}{(E-\epsilon)^2}
 \label{dos-eq}
\end{eqnarray}

The above expressions get enormously simplified as we set the resonance conditions $x^2+y^2= t^2$ and $\Phi=\frac{1}{4}\Phi_0$, and it is seen that under these conditions both $\rho_1$ and $\rho_3$ become identical to $\rho_C$ given above. This establishes the fact that the bands of the individual chains depicted in Fig.~\ref{decoupled}(a), (b), and (c) merge under the conditions for which the block-transfer matrices commute.


\section{}
The inverse participation ratio (IPR) is used to understand the localization properties of the eigenstates. It is defined as the fourth power of the normalized wave function~\cite{izrailev,tong}.
\begin{equation}
    IPR = \sum_n |\psi_n|^4
    \label{IPR-eq}
\end{equation}
where $n$ runs over all atomic sites. It gives us an understanding related to the distribution of the amplitudes of the wave function over a number of lattice sites. For example, for a completely localized eigenstate, the IPR tends to unity, whereas it is very close to zero for a perfectly extended eigenstate.



\begin{thebibliography}{99}
\bibitem{anderson} P. W. Anderson,  Absence of Diffusion in Certain Random Lattices, Phys. Rev. \textbf{109}, 1492 (1958).
\bibitem{tvr} P. A.  Lee and T. V. Ramakrishnan, Disordered electronic systems, Rev. Mod. Phys. \textbf{57}, 287 (1985).
\bibitem{abrahams} E. Abrahams, P. W. Anderson, D. C. Licciardello, and T. V. Ramakrishnan, Scaling Theory of Localization: Absence of Quantum Diffusion in Two Dimensions, Phys. Rev. Lett. \textbf{42}, 673 (1979).
\bibitem{borland} R. E. Borland, The nature of the electronic states in disordered one-dimensional systems, Proc. R. Soc. Lond. A \textbf{274}, 274529 (1963).
\bibitem{deylon} F. Deylon, Y. Le\'{e}vy, and B. Souillard, Anderson localization for one - and quasi one-dimensional systems, J. Stat. Phys. \textbf{41}, 375 (1985).
\bibitem{kramer} B. Kramer and A. MacKinnon, Localization: theory and experiment, Rep. Prog. Phys. \textbf{56}, 1469 (1993).
\bibitem{lagen} A. Lagendijk, B. van Tiggelen, and D. S. Wiersma, Fifty years of Anderson localization, Phys. Today \textbf{62}, 24, (2008).
\bibitem{albada} M. P. Van Albada and A. Lagendijk, Observation of weak localization of light in a random medium. Phys. Rev. Lett. \textbf{55}, 2692 (1985).
\bibitem{wiersma} D. S. Wiersma, P. Bartolini, A. Lagendijk, and R. Righini, Localization of light in a disordered medium. Nature \textbf{390}, 671 (1997).
\bibitem{martin} L. Martin, G. Di Giuseppe, A. Perez-Leija, R. Keil, F. Dreisow, M. Heinrich, S. Nolte, A. Szameit, A. F. Abouraddy, D. N. Christodoulides, and B. E. A. Saleh, Opt. Exp. \textbf{19}, 13636 (2011).
\bibitem{segev} M. Segev, Y. Silberberg, and D. N. Christodoulides, 
Anderson localization of light, Nat. Photonics,
\textbf{7}, 197 (2013).
\bibitem{roati} G. Roati, C. D’Errico1, L. Fallani, M. Fattori, C.  Fort1, M. Zaccanti1, G. Modugno1, M. Modugno, and M. Inguscio, Anderson localization of a non-interacting Bose–Einstein condensate, Nature \textbf{453}, 895 (2008). 
\bibitem{white} D. H. White, T. A. Haase, D. J. Brown, M. D. Hoogerland, M. S. Najafabadi, J. L. Helm, C. Gies, D. Schumayer, and D. A. W. Hutchinson, Observation of two-dimensional Anderson localisation of ultracold atoms, Nat. Comm. \textbf{11}, Article number: 4942 (2020). 
\bibitem{economou1} E. N. Economou and M. H. Cohen, Localization in Disordered Materials: Existence of Mobility Edges, Phys. Rev. Lett. \textbf{25}, 1445 (1970).
\bibitem{herbert} D. C. Herbert, and R. Jones, Localized states in disordered systems, J. Phys. C \textbf{4}, 1145 (1971).
\bibitem{thouless} D. J. Thouless, (1972). A relation between the density of states and range of localization for one dimensional random systems, J. Phys C \textbf{5}, 77 (1972).
\bibitem{economou2} D. Licciardello and E. N. Economou, A new criterion for localization in Anderson model for disordered lattices, Sol. State Comm. \textbf{15}, 1969 (1974).
\bibitem{anderson2} P. W. Anderson, D. J. Thouless, E. Abrahams, and D. S. Fisher, New method for a scaling theory of localization. Physical Review B, \textbf{22}, 3519 (1980).
\bibitem{guinea} F. Guinea and J. A. Vergés, Localization and topological disorder, Phys. Rev. B \textbf{35}, 979 (1987).
\bibitem{rudo0} R. A. R\"{o}mer, A. MacKinnon, and C. Villagonzalo, J. Phys. Soc. Jpn. \textbf{72} Suppl. A, 167 (2003).
\bibitem{francisco} F. Dominguez-Adame and V. A. Malyshev, A simple approach to Anderson localization in one-dimensional
disordered lattices, Am. J. Phys. \textbf{72}, 226 (2004).
\bibitem{sacha} K. Sacha, C. A. M\"{u}ller, D. Delande,  and J. Zakrzewski, Anderson Localization of Solitons, Phys. Rev. Lett. \textbf{103}, 210402 (2009).

\bibitem{sutherland} B. Sutherland, Localization of electronic wave functions due to local topology, Phys. Rev. B \textbf{34}, 5208 (1986).
\bibitem{bodyfelt}  D. Leykam, J. D. Bodyfelt, A. S. Desyatnikov, and S.
Flach, Localization of weakly disordered flat band states, Euro. Phys. J. B \textbf{90}, 1 (2017).
\bibitem{ajith} A. Ramachandran, A. Andreanov, and S. Flach, Chiral flat bands: Existence, engineering, and stability, Phys. Rev. B \textbf{96}, 161104(R) (2017).
\bibitem{sergej1} D. Leykam, A. Andreanov, and S. Flach, Artificial flat band systems: from lattice models to
experiments, Adv. in Phys.: X \textbf{3}, 1473052 (2018). DOI:
10.1080/23746149.2018.1473052
\bibitem{biplab} A. Bhattacharya and B. Pal, Flat bands and nontrivial topological properties in an extended Lieb lattice, Physical Review B \textbf{100}, 235145 (2019).
\bibitem{sergej2} C. Danieli, A. Andreanov, T. Mithun, and S. Flach, Nonlinear caging in all-bands-flat lattices, Phys. Rev. B \textbf{104}, 085131 (2021).
\bibitem{rudo1} J. Liu, X. Mao, J. Zhang, and R. A. R\"{o}mer, Localization properties in Lieb lattices and their extension, Annals of Phys. \textbf{435}, 168544 (2021).
\bibitem{rudo2} X. Mao, J. Liu, J. Zhong, and R. A. R\"{o}mer, Disorder effects in two dimensional Lieb lattice and its extension, Physica E \textbf{124}, 114340 (2020)
\bibitem{rudo3} J. Liu, X. Mao, J. Zhong, and R. A. R\"{o}mer, Localization, phases, and transitions in three-dimensional extended Lieb latticesPhys. Rev. B \textbf{102}, 174207 (2020).
.


\bibitem{dunlap} D. H. Dunlap, H-L. Wu, and Philip W. Phillips, Absence of localization in a random-dimer model, Phys. Rev. Lett. \textbf{65}, 88 (1990).
\bibitem{wu} H.-L. Wu and Philip Phillips, 194 citations
Polyaniline is a random-dimer model: A new transport mechanism for conducting polymers, Phys. Rev. Lett. \textbf{66}, 1366 (1991). 
\bibitem{goff} H.-L. Wu, William Goff, and Philip PhillipsInsulator-metal transitions in random lattices containing symmetrical defects, Phys. Rev. B \textbf{45}, 1623 (1992). 
\bibitem{kundu} P. K. Datta, D. Giri, and K. Kundu, Nonscattered states in a random-dimer model, Phys. Rev. B \textbf{47}, 10727 (1993).
\bibitem{sanchez} Angel Sánchez, Enrique Maciá, and Francisco Domínguez-Adame, Suppression of localization in Kronig-Penney models with correlated disorder, Phys. Rev. B \textbf{49}, 147 (1994).
\bibitem{heinrichs} J. Heinrichs, Localization, antilocalization, and delocalization in one-dimensional disordered lattices, Phys. Rev. B \textbf{51}, 5699 (1995).
\bibitem{grosso} R. Farchioni and G. Grosso, Electronic transport for random dimer-trimer model Hamiltonians, Phys. Rev. B \textbf{56}, 1170 (1997).
\bibitem{arunava1} S. Sil, S. N. Karmakar, R. K. Moitra, and A.  Chakrabarti, Extended states in one-dimensional lattices: Application to the quasiperiodic copper-mean chain, Phys. Rev. B \textbf{48}, 4192(R) (1993).
\bibitem{arunava2} A. Chakrabarti, S. N. Karmakar, and R. K. Moitra, Renormalization-group analysis of extended electronic states in one-dimensional quasiperiodic lattices, Phys. Rev. B \textbf{50}, 13276 (1994).
\bibitem{arunava3} A. Chakrabarti, S. N. Karmakar, and R. K. Moitra, Role of a New Type of Correlated Disorder in Extended Electronic States in the Thue-Morse Lattice, Phys. Rev. Lett. \textbf{74}, 1403 (1995).
\bibitem{macia1} E. Maciá and F. Domínguez-Adame, Physical Nature of Critical Wave Functions in Fibonacci Systems, Phys. Rev. Lett. \textbf{76}, 2957 (1996).
\bibitem{arunava4} A. Chakrabarti and B. Bhattacharyya, Atypical extended electronic states in an infinite Vicsek fractal: An exact result, Phys. Rev. B \textbf{54}, R12625(R) (1996).
\bibitem{wang} X. R. Wang, Phys. Rev. B \textbf{51}, 9310 (1995). 
\bibitem{arunava5} A. Chakrabarti, Exact results for infinite and finite Sierpinski gasket fractals: extended electron states and transmission properties, J. Phys.: Condens. Matter \textbf{8}, 10951 (1996).
\bibitem{arunava6} A. Chakrabarti, Extended electron states and magnetotransport in a 3-simplex fractal, Phys. Rev. B \textbf{72}, 134207 (2005).
\bibitem{skm} B. Pal, S. K. Maiti, and A. Chakrabarti, Complete absence of localization in a family of disordered lattices, Europhys. Lett. \textbf{102}, 17004 (2013).
\bibitem{bpac} B. Pal and A. Chakrabarti, Engineering bands of extended electronic states in a class of topologically disordered and quasiperiodic lattices, Phys. Lett. A \textbf{378}, 2782 (2014).
\bibitem{anbpac} A. Nandy, B. Pal, and A. Chakrabarti, Tight binding chains with off-diagonal disorder: bands of extended electronic states induced by minimal quasi-one dimensionality, Europhys. Lett. \textbf{115}, 37004 (2016).
\bibitem{amrita} A. Mukherjee, A. Chakrabarti, and R. A. R\"{o}mer. Flux driven and geometry controlled spin filtering for arbitrary spins in aperiodic quantum networks, Phys. Rev. B \textbf{98}, 075415 (2018).
\bibitem{arka} A. Maity and A. Chakrabarti, Engineering insulator-metal transition in a class of decorated lattices: a quantum dynamical study, Phys. Lett. A \textbf{406}, 127452 (2021).

\bibitem{andrade1} R. F. S. Andrade and H. J. Schellnhuber, Electronic states on a fractal: Exact Green’s-function renormalization approach, Phys. Rev. B \textbf{44}, 13213 (1991).
\bibitem{macia} E. Macia, Electronic transport in the Koch fractal lattice, Phys. rev. B \textbf{57}, 7661 (1998).
\bibitem{andrade2} R. F. S. Andrade and H. J. Schellnhuber, Electronic states on a fractal: The consequences of self-energy variation, Phys. Rev. B \textbf{55}, 12956 (1997).
P. Kappertz, R. F. S. Andrade, and H. J. Schellnhuber, Electronic states on a fractal: Inverse-iteration method, Phys. Rev. B \textbf{49}, 14711 (1994).
\bibitem{zheng} D. Zheng, Y. Liu, and  Z.D. Wang, Quantum conductivity exponent of a fractal non-branching Koch curve, Solid St. Commun. \textbf{98}, 527 (1996).
\bibitem{zhao} P. Zhao, K. Zhang and Z. Deng,  Elastic Wave Propagation in Lattice Metamaterials with Koch Fractal, Acta Mech. Solida Sinica, \textbf{33}, 600 (2020). 
\bibitem{achiam} Y. Achiam, Critical dynamics of the kinetic Ising model on the fractal Koch curves, Phys. Rev. B \textbf{32}, 1796 (1985).
\bibitem{zhang} Z. Zhang, S. Zhou, W. Xie, L. Chen, Y. Lin, and J. Guan, Standard random walks and trapping on the Koch network with scale-free behavior and small-world effect, Phys. Rev. E \textbf{79}, 061113 (2009).
\bibitem{arunava-koch1} A. Chakrabarti, Field induced delocalization in a Koch fractal, Phys. Rev. B \textbf{60}, 10576 (1999). 
\bibitem{decimation} Decimation means, eliminating the amplitudes $\psi_j$ corresponding to the vertices $A$ throughout the lattice in terms of the amplitudes at the sites $B$, $C$ and $D$. This is done by using the difference equation.
\bibitem{roman} H. E. Roman, Electronic properties of a one-dimensional hierarchical system, Phys. Rev. B \textbf{36}, 7173 (1987).
\bibitem{ceccato} 
H. A. Ceccatto, W. P. Keirstead, and B. A. Huberman, 
Quantum states of hierarchical systems, Phys. Rev. A \textbf{36}, 5509(R) (1987).
\bibitem{kunz} H. Kunz, R. Livi, and A. S\"{u}t\H{o}, Cantor spectrum and singular continuity for a hierarchical Hamiltonian, Communications in Mathematical Physics volume 122, pages643–679, (1989)

\bibitem{southern} B. W. Southern, A. A. Kumar, P. D. Loly, and A.-M. S. Tremblay, Real-space rescaling method for the spectral properties of tight-binding systems, Phys. Rev. B \textbf{27}, 1405(R) (1983).
\bibitem{arunava-koch2} A. Chakrabarti, Strange eigenstates and anomalous transport in a Koch fractal with hierarchical
interactions, Phys. Lett. A \textbf{375}, 3899(2011).
\bibitem{fisher} D. S. Fisher and P. A. Lee, Phys. Rev. B \textbf{23}, 6851 (1981).
\bibitem{supriya} S. datta, {\it Electronic Transport in Mesoscopic Systems}, Cambridge University Press, Cambridge (1997).
\bibitem{izrailev} F. M. Izrailev, A. A. Krokhin, and N. M. Makarov, Anomalous localization in low dimensional systems with correlated disorder, Phys. Rep. \textbf{512}, 125 (2012).
\bibitem{tong} P. Tong, B. Li, and B. Hu, Electronic properties of the 1D Frenkel-Kontorova model, Phys. Rev. Lett. \textbf{88}, 046804 (2002).

\end{thebibliography}
\end{document}